\documentclass[11pt]{iopart}
\usepackage{graphicx}
\usepackage{amssymb}
\usepackage{amsfonts}
\usepackage{dsfont} 
\usepackage{mathrsfs}
\usepackage{bold-extra}
\usepackage{relsize}
\usepackage{cite}
\usepackage{hyperref}
\usepackage{tensor}
\usepackage[cal=boondoxo]{mathalfa}

\usepackage{txfonts} 

\newcommand{\etat}{\tilde{\eta}}

\newcommand{\m}{\mu}
\newcommand{\w}{\omega}
\newcommand{\n}{\nu}
\newcommand{\s}{\sigma}
\newcommand{\nid}{\noindent}
\newcommand{\ph}[1]{\phantom{#1}}
\newcommand{\ovg}[1]{\stackrel{(\gamma)}{#1}}
\renewcommand{\ms}[1]{\mathcal{#1}}
\renewcommand{\e}{\hspace*{0.1pt}}
\newcommand{\mc}[1]{\mathcal{#1}}

\hypersetup{colorlinks=true,linkcolor=blue,urlcolor=blue,citecolor=blue}

\begin{document}

\title{Barbero's formulation from a $BF$-type action with the Immirzi parameter}

\author{Mariano Celada, Merced Montesinos and Jorge Romero}

\address{Departamento de F\'{i}sica, Cinvestav, Av. Instituto Polit\'{e}cnico Nacional 2508,
San Pedro Zacatenco, 07360, Gustavo A. Madero, Ciudad de M\'{e}xico, M\'{e}xico}

\ead{mcelada@fis.cinvestav.mx, merced@fis.cinvestav.mx, ljromero@fis.cinvestav.mx}

\begin{abstract}
Starting from a constrained real $BF$-type action for general relativity that includes both the Immirzi parameter and the cosmological constant, we obtain the Ashtekar-Barbero variables used in the canonical approach to the quantization of the gravitational field. This is accomplished by explicitly solving the second-class constraints resulting from the Hamiltonian analysis of the considered action, and later imposing the time gauge. All throughout this work the tetrad formalism is left aside, obtaining the Ashtekar-Barbero variables entirely in terms of the $B$-fields that define the action.
\end{abstract}

\vspace{2pc}
\noindent{\it Keywords}: BF gravity, Canonical general relativity, Immirzi parameter

\section{Introduction}
\label{SIntro}

The achievement of a consistent quantum theory of gravity has been one of the most tackled problems through the years and still remains as one of the most exciting open issues in theoretical physics. Amongst all the competing candidates, those which intend a non-perturbative and background-independent quantization of gravity have played a major role in the last two decades, being loop quantum gravity (LQG) \cite{lrrRovelli, Rovelli, Thiemann} and its covariant version, the spin foam models \cite{lrr-Perez}, two of the main approaches. LQG follows the canonical quantization program, and describes the quantum spatial geometry of general relativity. On the other hand, the spin foam approach, which intends a covariant path integral quantization of the gravitational field, supplements the loops approach and allows us to explore the full quantum dynamics of general relativity. 

The first hints for a viable canonical quantization of gravity were established when the celebrated ADM formulation \cite{ADM} came to light, although some technical and conceptual problems forced people to abandon this approach \cite{RovWdW}. This way of thinking was later revived when Ashtekar \cite{Ashtekar}, through a canonical transformation on the ADM variables, found a new set of complex variables for general relativity sharing many similarities with those used to describe the phase space of Yang-Mills theory. In terms of these variables, the constraints of general relativity became considerably simplified, and although a lot of progress was made using this new formalism, the reality conditions (necessary to recover real general relativity) were too difficult to handle at the quantum level.

The struggle with the reality conditions stopped when Barbero \cite{Barbero}, inspired by Ashtekar's ideas, discovered a new set of real variables for describing the phase space of general relativity. This formulation thus gave birth to LQG as we know it, even though the theory paid the price of having a more complicated scalar constraint to deal with. Later, Holst \cite{Holst} proposed an action principle from which the now called Ashtekar-Barbero variables could be derived. This action, however, contains a free parameter (the Immirzi parameter \cite{Immirzi}) which has no importance at the classical level, but whose meaning at the quantum level still remains unclear since the spectra of geometric operators \cite{Lewandowski,Smolin} as well as the entropy of black holes \cite{RovelliHole,ABK,Meis,Agul,JKA} explicitly depend on it. However, some newer results indicate that by choosing the connection appropriately one can get rid of this parameter from the quantum theory  \cite{Alexandrov,Alexandrov1,Alexandrov2,Alexandrov3,Karim,Geiller,Karim2}.

On the other hand, the spin foam models (the covariant approach to LQG) use variations of the action principle first introduced by Plebanski \cite{Plebanski} in the mid 70's, who realized that (complex) general relativity could be formulated as a $BF$ theory (although the terminology $BF$ appeared later) with some constraints on the $B$-field necessary to recover the tetrad field. These models aim to achieving a path integral quantization of the gravitational field keeping its fundamental classical symmetries untouched (that is, without breaking neither the manifest diffeomorphism invariance nor the local Lorentz invariance of the theory, as most of the canonical methods do). Spin foams then become transition amplitudes between quantum states of 3-geometries (the spin-networks LQG uses to describe the quantum spatial geometry of gravity), and so they can be used for exploring the full quantum dynamics of general relativity, something that has been difficult to carry out in LQG since the implementation of the scalar constraint at the quantum level is quite complicated \cite{AshLew}. 

Although the $BF$-type actions are fundamental for the development of the spin foam models, they also can be used to understand other aspects of general relativity. Based on the action introduced by Plebanski, it was proved in Ref. \cite{Capovilla} that the Ashtekar formalism was the result of the canonical analysis of such an action, and this provided a much simpler path to obtain the Ashtekar variables than the one devised by Ashtekar. Following this line of thought, it must be possible to obtain the Ashtekar-Barbero variables from a real $BF$ formulation of general relativity that includes the Immirzi parameter. An action of this kind is reported in Ref. \cite{CMPR} (the CMPR action). The first part of the Hamiltonian analysis of this action (and of another related to it) is carried out in Ref. \cite{Celada}, and it is shown there that besides the usual first-class constraints intrinsic to general relativity, the theory possesses second-class constraints. Now, when dealing with second-class constraints, there are two equivalent methods that can be followed \cite{Teitel}, namely, either we can introduce the Dirac bracket, or we can explicitly solve the second-class constraints. In this paper we follow the latter and show that we recover the Ashtekar-Barbero variables after solving the second-class constraints and imposing a gauge fixing.

In order to arrive at Barbero's formulation, in section \ref{SHamiltonian} we present a summary of the Hamiltonian analysis reported in Ref. \cite{Celada} and establish the conventions that we follow. In section \ref{SSolution} we solve the second-class constraints and describe the phase space of general relativity in terms of the resulting variables. Finally, in section \ref{SGauge} we show how the Ashtekar-Barbero variables emerge when the time gauge is taken.

\section{Hamiltonian analysis of $BF$ gravity with cosmological constant}
\label{SHamiltonian}

Throughout this paper Greek letters denote tangent space indices, and Latin capital letters denote group indices that are lowered and raised with the internal metric $(\eta_{IJ})=\mbox{diag}(\sigma,1,1,1)$, where $\s=-1\ [\sigma=+1]$ for the gauge group SO(1,3) [SO(4)]. 


When the 3+1 decomposition is performed, the spacetime manifold $M$ is assumed to have a $\mbox{topology}$ $M\approx\mathbb{R}\times\Omega$, with $\Omega$ a compact 3-manifold without boundary. The indices ``$t$" and ``$0$" label the time component in the coordinate basis and in the internal space, respectively; on the other hand, the indices starting at ``$a$" label the spatial components of tensors in the coordinate basis, while those starting at ``$i$" do the same in the internal one.

General relativity with the Immirzi parameter and cosmological constant can be described by the following (real) $BF$-type action \cite{Velazquez,Velazquez2,MontesMer2}

\begin{eqnarray}
	\fl S[B,\omega,\phi,\mu]=\int_M&\left[\left(B^{IJ}+\frac{1}{\gamma}\ast B^{IJ}\right)\wedge F_{IJ}[\omega]-\phi_{IJKL}B^{IJ}\wedge B^{KL} -\mu\phi_{IJKL}\epsilon^{IJKL} \right. \nonumber\\
	&\hspace{5mm}\left. +\mu\lambda + l_{1}B_{IJ}\wedge B^{IJ} + l_{2} B_{IJ}\wedge\ast B^{IJ}\ph{\frac{1}{\gamma}} \right],\label{BFact}
\end{eqnarray}

\nid where $B^{IJ}$ is a set of six 2-forms ($B^{IJ}=-B^{JI}$), $F_{IJ}$ is the curvature for the connection $\omega_{IJ}$ and its components are given by	$F_{\m\n IJ}= 2(\ \ph{\frac{1}{2}}\hspace{-0.3cm}\partial_{[\m}\w_{\n] IJ} + \w_{[\m|I}\e^K \w_{|\n]KJ})$, the terms involving $\lambda$, $l_1$ and $l_2$ are added in order to incorporate the cosmological constant \cite{Velazquez2}, $\phi_{IJKL}$ is an internal tensor imposing certain constraints on the $B$-field which has the symmetries $\phi_{IJKL}=\phi_{KLIJ}=-\phi_{JIKL}=-\phi_{IJLK}$, $\m$ is a 4-form, and $\gamma$ is the Immirzi parameter. The symbol $\ast$ stands for the Hodge internal dual: $\ast U_{IJ}=(1/2) \epsilon_{IJKL}U^{KL}$ for $U_{IJ}=-U_{JI}$, and $\epsilon_{0123}=1$. We point out that the action (\ref{BFact}) involves the two $BF$ terms allowed by the two bilinear forms of the Lie algebras of SO(1,3) [or SO(4)] \cite{MontesBF_2}, but it is not the only action of the $BF$-type for general relativity that integrates the Immirzi parameter into the theory. In fact, the CMPR action \cite{CMPR}, which considers a linear combination of the two invariants $\epsilon_{IJKL}\phi^{IJKL}$ and $\tensor{\phi}{^{IJ}_{IJ}}$ but just one of the $BF$ terms, also does the job. Although we could also consider the CMPR action from the very beginning, the results obtained from the action (\ref{BFact}) also apply to the CMPR action, as sketched below (this is a consequence of the fact that both actions can be mapped into each other at the Lagrangian \cite{Velazquez} and Hamiltonian \cite{Celada} levels).

To simplify future computations it is useful to define:    

\begin{equation}
	\label{ovg}
	\ovg{U}_{IJ}:=U_{IJ}+\frac{1}{\gamma}\ast U_{IJ},
\end{equation}

\nid whose inverse transformation 

\begin{equation}
	U_{IJ}=\frac{\gamma^2}{\gamma^2-\sigma}\left(\ovg{U}_{IJ}-\frac{1}{\gamma}\ast \ovg{U}_{IJ}\right),
\end{equation}

\nid
holds as long as  $\gamma^{2}\neq\s$. These definitions provide the following identities

\begin{equation}
	\ovg{U}_{IJ}V^{IJ}=U_{IJ}\ovg{V}\tensor{}{^{IJ}},
\end{equation}

\begin{equation}
	\ovg{\left(\ph{\frac{1}{2}}\hspace{-0.3cm}U_{[I|}\e^K V_{K|J]}\right)}=\ovg{U}\tensor{}{_{[I|}^K} V_{K|J]} = \tensor{U}{_{[I|}^K} \ovg{V}\tensor{}{_{K|J]}},
\end{equation}
 
\begin{equation}
	\ast\left(\ph{\frac{1}{2}}\hspace{-0.3cm}U_{[I|}\e^K V_{K|J]}\right)=\ast U_{[I|}\e^K V_{K|J]} = U_{[I|}\e^K \ast V_{K|J]}.
\end{equation}

The first part of the Hamiltonian analysis of (\ref{BFact}) is reported in Ref. \cite{Celada}, where the authors find that the Hamiltonian action takes the form  

\begin{eqnarray}\label{bfcc2}
	\fl S[\omega_a,\mathop{\Pi}^{(\gamma)}{}^a,\ms{N},\ms{N}^a,\xi,\varphi_{ab},\psi_{ab}]=\int_\mathbb{R} dt\int_\Omega d^3x\ \Biggl(& \mathop{\Pi}^{(\gamma)}\tensor{}{^{aIJ}}\dot{\w}_{aIJ}+\ms{N}\ms{H} + \ms{N}^a\ms{H}_a + \xi_{IJ}\ms{G}^{IJ}\nonumber\\ 
	 &+\varphi_{ab}\Phi^{ab}+\psi_{ab}\Psi^{ab}\Biggr),
\end{eqnarray}

\nid where $\stackrel{(\gamma)}{\Pi}\tensor{}{^{aIJ}}$ are the canonically conjugate momenta to the spatial part of the connection $\w_{aIJ}$, and $\tensor{\Pi}{^{aIJ}}$ is defined in terms of the $B$-fields as

\begin{equation}
	\label{Pi}
	\Pi^{aIJ} := \frac{1}{2}\etat^{abc}B_{bc}\e^{IJ},  
\end{equation}

\nid
with $\etat^{abc}$ a totally antisymmetric tensor density of weight +1 ($\etat^{123}=1$); notice that the canonical pair can be equivalently chosen as $(\stackrel{(\gamma)}{\omega}\tensor{}{_{aIJ}}, \Pi^{aIJ})$, and it turns out that this form is more convenient to solve the second-class constraints. $\ms{N}$, $\ms{N}^{a}$, $\xi_{IJ}$, $\varphi_{ab}$, and $\psi_{ab}$ play the role of Lagrange multipliers imposing the following constraints



\begin{eqnarray}
 	\ms{G}^{IJ}&:=& D_a \ovg{\Pi}\tensor{}{^{aIJ}} \approx 0, \\
	\ms{H}_a &:=& \frac{1}{2}\Pi^{bIJ}\ovg{F}\tensor{}{_{baIJ}} \approx 0,\label{GaussLo} \\
	\ms{H}&:=& \frac{1}{4} \tilde{\eta}^{abc} h_{ad} \ast\Pi^{dIJ}\ovg{F}\tensor{}{_{bcIJ}} + 2\Lambda h \approx 0 ,\label{hamilcons}\\
	\Phi^{ab} & := & -\s\ast\Pi^{aIJ}\Pi^b\e_{IJ} \approx 0, \label{phi}\\
	\Psi^{ab} & := & 2h_{cf}\etat^{(a|cd}\Pi^f\e _{IJ}D_d \Pi^{|b)IJ}\approx 0\label{psi},   	
\end{eqnarray}


\nid
where $\Lambda=3l_2-\s\lambda/4$ is the cosmological constant, $D_a$ is the SO(1,3) [or SO(4)] covariant derivative ($D_a \Pi^{bIJ}=\partial_{a}\Pi^{bIJ} + \w_{a}\e^{I}\e_{K}\Pi^{bKJ}-\w_{a}\e^{J}\e_{K}\Pi^{bKI}$), and $h$ is the determinant of the spatial metric\footnote{Notice that in the $BF$-type actions such as (\ref{BFact}) there is no spacetime metric a priori. The fundamental variables in this formulation are the $B$-fields instead, and the spacetime metric is constructed from them \cite{CMPR}, thus becoming a derived object. Here we are assuming that $h_{ab}$ corresponds to a metric on the spatial 3-manifold $\Omega$, something that is made clear when we write the $B$-fields in terms of the tetrad field. In fact, since in a diffeomorphism-invariant theory with a metric the spatial metric always appears as a structure function in the Poisson bracket of the scalar constraint with itself \cite{Hojman, Peldan1, Peldan2}, and since the same role is played by $h_{ab}$ for the scalar constraint (\ref{hamilcons}) (see \cite{Celada}), then this supports our assumption that $h_{ab}$ is to be interpreted as the spatial metric on the hypersurface.} $h_{ab}$, whose inverse $h^{ab}$ satisfies the relation

\begin{equation}
	\label{h}
	hh^{ab}=\frac{\s}{2} \Pi^{aIJ}\Pi^{b}\e_{IJ}.
\end{equation}

The classification of the constraints is as follows. $\ms{G}^{IJ}$, $\ms{H}_a$, and $\ms{H}$ are first class (they are known as Gauss, vector, and scalar constraints, respectively) and thereby generate the gauge transformations of the theory, namely, the former generates local Lorentz transformations (rotations in the case of Euclidean signature), while the other two are related to the generator of spacetime diffeomorphisms. On the other side, $\Phi^{ab}$ and $\Psi^{ab}$ are second-class constraints. Hence, the counting of the physical degrees of freedom is straightforward and yields two per space point, the number any description of Einstein's gravity must have.

Although the canonical analysis presented above does not involve the tetrad field, it is possible to recover the tetrad by solving the constraints imposed on the $B$-fields by the Lagrange multiplier $\phi_{IJKL}$  in (\ref{BFact}). By doing this, the $B$-fields take the form $B^{IJ}=\ast\left(e^{I}\wedge e^{J}\right)$, and this expression redefines $\Pi^{aIJ}$ [$\Pi^{aIJ}=2e e^{t[I|}e^{a|J]}$, where $e=\det (e_{\m}\e^{I})$] as well as all the constraints involved; the resultant canonical analysis turns out to be compatible (for $\s=-1$ and $\Lambda=0$), up to a multiplicative constant for each constraint, with the one carried out starting from the Holst action \cite{Barros}. Even though it is possible to continue this path and solve the second-class constraints in terms of the tetrad fields to reach the Ashtekar-Barbero variables, this work proceeds directly from the variables defined in terms of the $B$-fields.

\section{Solving the second-class constraints}\label{SSolution}

According to Dirac's method, second-class constraints can be used to eliminate some canonical variables from the formalism. Therefore, the solution to the equations (\ref{phi}) and (\ref{psi}) reduces the original phase space with 36 canonical variables $(\w_{aIJ},\stackrel{(\gamma)}{\Pi}\tensor{}{^{aIJ}})$ [or $(\ovg{\w}_{aIJ},\Pi^{aIJ})$] to a simpler one with 24 canonical variables; more explicitly, the constraints (\ref{phi}) and (\ref{psi}) allows us to eliminate 6 of the variables $\stackrel{(\gamma)}{\Pi}\tensor{}{^{aIJ}}$ and 6 of the variables $\w_{aIJ}$, respectively. This done as follows. First, the solution to (\ref{phi}) is

\begin{eqnarray}
	\Pi^{a0i}&=&E^{ai},  \label{sol1a}\\
	\Pi^{aij}&=&2E^{a[i}\chi^{j]}. \label{sol1b}
\end{eqnarray}

\nid\nid We assume that $E^{ai}$ is invertible with inverse $E_{ai}$ \footnote{$E^{ai}$ and $\chi^i$ thus constitute a parametrization of the original canonical variables $\Pi^{aIJ}$ such that the constraint $\Phi^{ab}$ is solved. In terms of $\Pi^{aIJ}$ they read	\begin{eqnarray*}
		E^{ai} &=&\Pi^{a0i}, \\
		\chi^{i} &=&-\frac{1}{2}E_{aj}\Pi^{aij},
\end{eqnarray*}
	
\nid that is, the independent variables are the nine ``electric'' components $\Pi^{a0i}$ plus three of the ``magnetic'' components $\Pi^{aij}$.
}. This solution redefines (\ref{h}) as:

\begin{equation}
	\label{hh}
	hh^{ab}=\eta_{ij}E^{ai}E^{bj},
\end{equation}

\nid
where $\eta_{ij}$ is given by

\begin{equation}\label{eta}
	\eta_{ij}: =\left(1+\s\chi_{k} \chi^{k} \right)\delta_{ij}-\s\chi_{i}\chi_{j},
\end{equation}
and the spatial internal indices are raised and lowered with the Euclidean metric $\delta_{ij}$.

On the other hand, notice that using (\ref{sol1a}) and (\ref{sol1b}), the symplectic structure collapses to 

\begin{equation}\label{reducedpsv}
	\hspace{-15mm}\int_{\Omega}d^{3}x\left( \mathop{\Pi}^{(\gamma)}{^{aIJ}}\dot{\w}_{aIJ}\right)=\int_{\Omega}d^{3}x\left(\Pi^{aIJ}\frac{\partial}{\partial t}\ovg{\w}_{aIJ}\right) = 2\int_{\Omega}d^{3}x\left( E^{ai}\dot{A}_{ai} +  \zeta_i \dot{\chi}^i\right),
\end{equation}

\nid where we have defined the nine plus three variables 

\begin{eqnarray}
	\label{A}
	A_{ai} & := & \ovg{\w}_{a0i} + \ovg{\w}_{aij}\chi^{j}, \\
	\label{z}
	\zeta_i & := & \ovg{\w}_{aij}E^{aj}.
\end{eqnarray}

\nid According to (\ref{reducedpsv}), the phase space variables are now the pairs $\left(A_{ai},E^{ai} \right)$ and $\left( \chi^{i}, \zeta_{i} \right)$, which obey the Poisson brackets:

\begin{eqnarray}
	\left\lbrace A_{ai}(x),E^{bj}(y) \right\rbrace & = & \frac{1}{2}\delta^{b}_{a}\delta^{j}_{i}\delta^{3}(x,y),\\
	\left\lbrace \chi^{i}(x),\zeta_{j}(y) \right\rbrace & = & \frac{1}{2}\delta^{i}_{j}\delta^{3}(x,y).
\end{eqnarray}

\nid Equation (\ref{z}) can be thought of as an inhomogeneous linear system of equations for the unknowns $\ovg{\w}_{aij}$, which has the general solution 

\begin{equation}\label{wij}
	\ovg{\w}_{aij}= \frac{1}{2}\epsilon_{ijk}E_{al}M^{kl} - E_{a[i}\zeta_{j]},
\end{equation}

\nid where $M^{ij}=M^{ji}$ are six arbitrary variables involved in the general solution of the homogeneous linear system of equations associated to (\ref{z}). From (\ref{A}), we get 
\begin{equation}\label{w0j}
	\ovg{\w}_{a0i} = A_{ai} - \ovg{\w}_{aij}\chi^{j}.
\end{equation}

\nid Therefore, (\ref{wij}) and (\ref{w0j}) establish a linear map from $(A_{ai}, \zeta_i, M^{ij})$ into $(\ovg{\w}_{a0i}, \ovg{\w}_{aij})$. The inverse map, which maps $(\ovg{\w}_{a0i}, \ovg{\w}_{aij})$ into $(A_{ai}, \zeta_i, M^{ij})$, is given by (\ref{A}), (\ref{z}), together with 

\begin{eqnarray}
M^{ij} = E^{a(i}\epsilon^{j)kl}\ovg{\w}_{akl}.
\end{eqnarray}

In order to be consistent with the results of the canonical analysis of the Holst action \cite{Barros}, the form of $M^{ij}$ is proposed as:

\begin{eqnarray}\label{M}
	M^{ij} = & \frac{1}{1+\s \chi_{m}\chi^{m}}\left[ \left(f^k\e_k+\s f_{kl}\chi^k\chi^l \right)\delta^{ij} + \left(\s f^k\e_k - f_{kl}\chi^k\chi^l \right)\chi^i\chi^j \right. \nonumber \\
	& -f^{ij} - f^{ji} - \left. \s\left(f^{ik}\chi^j + f^{jk}\chi^i + f^{ki}\chi^j + f^{kj}\chi^i \right)\chi_k \right].
\end{eqnarray}

\nid
Substituting (\ref{sol1a}), (\ref{sol1b}), (\ref{wij}), (\ref{w0j}), and (\ref{M}) into (\ref{psi}), $\Psi^{ab}$ acquires the form of a linear system of equations for $f_{ij}$:

\begin{eqnarray}
\Psi^{ab}=T^{abij}f_{ij}+G^{ab}\approx 0,
\end{eqnarray}

\nid
where $T^{abij}$ and $G^{ab}$ depend on $E^{ai}$, $A_{ai}$, $\chi^{i}$ and $\zeta_{i}$ (and some of their derivatives). The explicit forms of $T^{abij}$ and $G^{ab}$ as well as the inversion of $T^{abij}$ are long and tedious, and so only the final result is presented:

\begin{eqnarray}
	f_{ij} & = & -\epsilon_{ikl}E^{ak}\left[\left(1-\s\gamma^{-2}\right) E_{bj} \partial_a E^{bl} + \s\chi^l A_{aj} \right] \nonumber \\ 
	& & +\frac{\s}{\gamma}\left(E^{ak}A_{ak}\delta_{ij}-A_{ai}E^{a}\e_j+\zeta_i\chi_j \right).\label{f}
\end{eqnarray}

Thus, because of (\ref{M}) and (\ref{f}), solving the second-class constraints (\ref{psi}) amounts to fixing $M^{ij}$. By substituting this expression for $M^{ij}$ into the right-hand side of (\ref{wij}) and (\ref{w0j}), $(\ovg{\w}_{a0i}, \ovg{\w}_{aij})$ depend on the reduced phase space variables $(E^{ai}, \chi^i, A_{ai}, \zeta_i)$ only. In summary, the  original set of canonical variables $(\w_{aIJ},\stackrel{(\gamma)}{\Pi}\tensor{}{^{aIJ}})$ [or $(\ovg{\w}_{aIJ},\Pi^{aIJ})$] collapses to $(E^{ai}, \chi^i, A_{ai}, \zeta_i)$ once the second-class constraints (\ref{phi}) and (\ref{psi}) are solved.

With this, it is now possible to rewrite the remaining constraints appearing in (\ref{bfcc2}) in terms of the new phase space variables, the result being a theory with only first-class constraints. The Gauss constraint is split into its boost and rotational parts as follows:

\begin{eqnarray}
	\hspace{-10mm}\ms{G}^i_{\mbox{{\scriptsize boost}}}& :=  \ms{G}^{0i}  =  \partial_a\left(E^{ai}+\frac{\s}{\gamma}\epsilon^i\e_{jk}E^{aj} \chi^k  \right) +2\s A_{aj} E^{a[j}\chi^{i]}+\s\zeta_j\chi^j\chi^i+\zeta^i, \\
	\hspace{-10mm}\ms{G}^i_{\mbox{{\scriptsize rot}}} := & \frac{1}{2}\epsilon^{i}\e_{jk}\ms{G}^{jk} =  \partial_a\left(\epsilon^i\e_{jk}E^{aj}\chi^k  + \frac{1}{\gamma}E^{ai} \right) - \epsilon^i\e_{jk}\left(A_a\e^jE^{ak} -\zeta^j\chi^k\right).  
\end{eqnarray}

Using the Gauss constraint, the vector and scalar constraints take the form
\begin{eqnarray}
	\ms{H}_a & = & 2E^{bi}\partial_{[b}A_{a]i}-\zeta_i\partial_a \chi^i +\frac{\gamma^2}{\gamma^2-\s}\left[\ph{\frac{1}{2}}\hspace{-0.3cm} 2\s E^{b[i}\chi^{j]}A_{ai}A_{bj} \right. \nonumber\\
	& &  \left. - A_{ai}\left( \zeta^i +\s \zeta_j\chi^j\chi^i\right) + \frac{\s}{\gamma}\epsilon_{ijk}\left( E^{bi}A_b\e^j+\zeta^i\chi^j \right)A_a\e^k  \right],
\end{eqnarray}

\begin{eqnarray}
	&\fl\ms{H} =  -\s E^{ai}\chi_i \ms{H}_a + \left(1+\s\chi_{k} \chi^{k} \right) \left( E^{ai}\partial_a\zeta_i + \frac{1}{2}\zeta_i E^{ai}E^{bj}\partial_aE_{bj} \right) \nonumber \\
	 & \hspace{-16mm}-\frac{\s\gamma^2}{\gamma^2-\s}\left\lbrace \left(1+\s\chi_{l} \chi^{l} \right) \left[ E^{ai}E^{bj}A_{a[i|}A_{b|j]} + \zeta_i\chi^i A_{aj}E^{aj} + \frac{3}{4}\left(\zeta_i \chi^i\right)^2 + \frac{3}{4}\sigma\zeta_i \zeta^i  \right.\right. \nonumber \\
	 &  \hspace{-16mm}\left. - \frac{1}{\gamma}\epsilon_{ijk}E^{ai} A_{a}\e^j \zeta^k \right] -\frac{\s}{4}\left(f^i\e_i\right)^2 +\frac{\s}{2}f^{ij}f_{(ij)}  - \frac{1}{2}f_{ij}\chi^i\chi^j \left(f^k\e_k-\frac{\s}{2} f_{kl}\chi^k\chi^l\right)  \nonumber \\
	 & \hspace{-16mm}\left.  + \frac{1}{4}\left(f_{ij}+f_{ji} \right)\left(f^{ik}+f^{ki} \right)\chi^j\chi_k  \right\rbrace +2\Lambda\left(1+\s\chi_{k} \chi^{k} \right) E,\label{scalarun}
\end{eqnarray}

\nid where $E:=\det (E^{ai})$ is related to $h$ through $h^2=\left(1+\s\chi_{k} \chi^{k} \right)^2 E^2$. Notice that in spite of us having solved the second-class constraints, the number of physical degrees of freedom has not been modified. Our phase space is now described by 12 canonical pairs $\{(A_{ai},E^{ai}),(\chi^i,\zeta_i)\}$  subject to the ten first-class constraints $\ms{G}^i_{\mbox{{\scriptsize boost}}}$, $\ms{G}^i_{\mbox{{\scriptsize rot}}}$, $\ms{H}_a$, and $\ms{H}$; thus leaving a phase space with the same two physical degrees of freedom per space point mentioned in section \ref{SHamiltonian}.

To reach Barbero's formulation we demand the variables $E^{ai}$ to constitute a (densitized) triad for the spatial metric $h_{ab}$, and this requirement is fulfilled when $\chi^{i}=0$ [see Eqs. (\ref{hh}) and (\ref{eta})], which is also known as time gauge. Although a direct computation from the previous results should lead to Barbero's formulation, this path is rather complicated and therefore, by following the guidance of Ref. \cite{Barros}, an easier way is presented in the next section.

\section{Gauge fixing}
\label{SGauge}

Until now, the analysis has the full gauge freedom, that is, the theory is fully diffeomorphism and Lorentz invariant [SO(4) invariant in the case of Euclidean signature]. In order to derive the Ashtekar-Barbero variables, it is necessary to break the Lorentz group SO(1,3) down to its compact subgroup SO(3). To this end, the boost arbitrariness must be gauge fixed [in the case of Euclidean signature, since $\mathfrak{so}(4)\approx \mathfrak{so}(3)\oplus \mathfrak{so}(3)$, we fix one of the two SO(3) subgroups], and this is accomplished by choosing the time gauge $\chi^{i}=0$. It is easier to fix the gauge before obtaining the solution to the second-class constraints, and so in this section we repeat the derivations of section \ref{SSolution} but assuming $\chi^{i}=0$ from the outset. The solution of the second-class constraints is then given by

\begin{eqnarray}
	\Pi^{a0i} &=& E^{ai}, \\
	\Pi^{aij} &=& 0,
\end{eqnarray}

\nid
whereas $\ovg{\w}_{a0i}$ and $\ovg{\w}_{aij}$ can be written as:
\begin{eqnarray}
	\ovg{\w}_{a0i} &=& A_{ai}, \\
	\ovg{\w}_{aij} &=& \epsilon_{ijk}\left[\left(1 - \s\gamma^{-2}\right)\Gamma_{a}\e^k +\s \gamma^{-1}A_a\e^k\right],
\end{eqnarray}

\nid
where $\Gamma_{ai}$ is the rotational part of $\w_{aIJ}$:

\begin{equation}
	\label{Gamma}
	\Gamma_{ai}:=\frac{1}{2}\epsilon_{ijk}\w_{a}\e^{jk}.
\end{equation}

\nid
These expressions allow us to rewrite the constraints  $\ms{G}^{IJ}$ and $\Psi^{ab}$ as follows:
\begin{eqnarray}
	\ms{G}^i_{\mbox{{\scriptsize boost}}} & = & \partial_a E^{ai}+\epsilon^{ijk}  \left[\left(1-\s\gamma^{-2}\right)\Gamma_{ak}+\s\gamma^{-1}A_{ak}\right]E^{a}\e_{j}, \label{fix1}\\
	\ms{G}^i_{\mbox{{\scriptsize rot}}} & = & \gamma^{-1}\partial_aE^{ai}-\epsilon^{ijk}A_{aj}E^a\e_k, \label{fix2}\\
	\Psi^{ab} &=& -4\s \epsilon_{ijk}E^{ci}E^{(a|k}\left(\ph{\frac{1}{2}}\hspace{-0.3cm}\partial_cE^{|b)j}-\epsilon^{jlm}\Gamma_{cl}E^{|b)}\e_m\right). \label{fix3}
\end{eqnarray}

\nid
The combination of (\ref{fix1}) and (\ref{fix2}) yields the equations

\begin{equation}
	\frac{\gamma^{2}}{\gamma^{2}-\sigma} \left(\mc{G}^i_{\mbox{{\scriptsize boost}}} - \frac{\sigma}{\gamma}\mc{G}^i_{\mbox{{\scriptsize rot}}}\right) = \partial_{a}E^{a i} + \epsilon^{ijk} \Gamma_{ak}E^{a}\e_{j}=0,
\end{equation}

\nid
which, together with (\ref{fix2}), provide a system of equations for $\Gamma_{ai}$ whose solution is

\begin{equation}
\Gamma_{ai}=\epsilon_{ijk}\left(\ph{\frac{1}{2}} \hspace{-0.3cm} \partial_{[b}E_{a]}\e^j+E_{a}\e^{[l|}E^{c|j]}\partial_bE_{cl}\right)E^{bk}.
\end{equation}

\nid
Hence, this connection turns out to be the spin-connection of the densitized triad $E^{ai}$. The remaining constrains can now be written in terms of the new phase space variables ($A_{ai},E^{ai}$), and their final expressions are (the rotational label has been dropped from the Gauss constraint) 
\begin{eqnarray}
	\ms{G}^i & = & \gamma^{-1}\mathfrak{D}_{a}E^{ai},\label{constg}\\
	\ms{H}_a & = & E^{bi}F_{bai} + \left( 1 -\s \gamma^{-2} \right)\left(\gamma A_{ai} - \Gamma_{ai}\right)\ms{G}^i, \label{consth}\\
	\ms{H} & = & \frac{\s}{2\gamma} \epsilon_{ijk} E^{aj} E^{bk} \left[ F_{ab}\e^i + \left( \s\gamma -  \gamma^{-1}\right) R_{ab}\e^i \right] + 2\Lambda E,\label{consthh}
\end{eqnarray}

\nid
where $\mathfrak{D}_{a}E^{ai}:=\partial_aE^{ai}-\gamma\epsilon^{ijk}A_{aj}E^a\e_k$, and $F_{abi}$ and $R_{abi}$ are the curvatures of the connections $A_{ai}$ and $\Gamma_{ai}$, respectively:
\begin{eqnarray}
	F_{abi} & = & 2\ \partial_{[a}A_{b]i}-\gamma\epsilon_{ijk}A_a\e^jA_b\e^k, \\
	R_{abi} & = & 2\ \partial_{[a}\Gamma_{b]i}-\epsilon_{ijk}\Gamma_a\e^j\Gamma_b\e^k.
\end{eqnarray}

The description of the phase space of general relativity in terms of the canonical pair $(A_{ai},E^{ai})$ subject to the first-class constraints $\ms{G}^i$, $\ms{H}_a$, and $\ms{H}$ constitutes Barbero's formulation (Barbero originally assumed that $\gamma=\pm1$, but the formulation is valid for $\gamma\in\mathbb{R}-\{0\}$). Although all the analysis has been restricted to the case $\gamma^{2}\neq\s$, the final result does not forbid this value, and then it is even possible to recover the Ashtekar formulation when $\s=-1$ and $\gamma=\pm i$.

\section{Conclusions}
\label{SConclusions}
 
In this paper, we have shown that the canonical analysis of the real $BF$-type action principle (\ref{BFact}) leads to a description of the phase space of (real) general relativity in terms of the Ashtekar-Barbero variables, which is in full agreement with the fact that the Ashtekar variables for complex general relativity can be derived from Plebanski's action \cite{Capovilla}. It is remarkable to mention that during the whole process the tetrad field was not involved.

We started from the results of Ref. \cite{Celada}, where the constraints arising from the action principle (\ref{BFact}) are derived in a Lorentz-covariant fashion. Since some of those constraints are second class, our strategy consisted in explicitly solving them. In this way, we obtain a phase space with the same local degrees of freedom as the original one described by fewer canonical variables subject to first-class constraints only. Note that the solution of the second-class constraints (\ref{phi}) and (\ref{psi}) (see section \ref{SSolution}) requires splitting the Lorentz group into rotations and boosts, but this does not affect the Lorentz covariance of the theory. Barbero's formulation is then recovered after imposing the time gauge, which kills the boost freedom [in the case of Euclidean signature, we kill one of the two SO(3) subgroups].

Our results agree with those of Ref. \cite{Barros}, where the canonical analysis of the Holst action is carried out. This is expected, since the constraints (\ref{GaussLo})-(\ref{psi}) agree with those of Ref. \cite{Barros} when the tetrad field is introduced. In Ref. \cite{Barros}, the author also solves the second-class constraints arising from his analysis and makes contact with Barbero's formulation. Nevertheless, there is one difference: the expression (\ref{scalarun}) for the scalar constraint has the same form of the expression he obtains (see equation (62) in Ref. \cite{Barros}), but some of the terms involving products of $f_{ij}$ with itself in (\ref{scalarun}) do not get multiplied by $1+\sigma\chi_i\chi^i$ but just by 1. This difference, however, disappears when the time gauge is taken, although it becomes relevant when exploring the full dynamics of the theory without taking any specific gauge. An alternative derivation of Barbero's formulation starting from a $BF$-type action is also reported in Ref. \cite{lrr-Perez}, but the time gauge is used there to solve the simplicity constraints on the $B$-fields, and so no Lorentz-covariant treatment is considered. 

Since our results agree with those of Ref. \cite{Barros}, and since this formulation is used to explore the role of the Immirzi parameter in Lorentz-covariant loop quantum gravity \cite{Karim,Geiller}, then a loop quantization of the theory contained in Sect. \ref{SSolution} should lead to the same quantum results. That is, that the role of the Immirzi parameter at the quantum level would depend on the connection used to construct the holonomies; in particular, by using the unique Lorentz-covariant connection found in \cite{Karim,Geiller}, the eigenvalues of the area operator are $\gamma$-dependent.

Finally, regarding the CMPR action, we mentioned in section \ref{SHamiltonian} that it and the action principle (\ref{BFact}) can be mapped into each other. From the Hamiltonian point of view, this is achieved via the transformation $\stackrel{(\gamma)}{\Pi}\tensor{}{^{aIJ}}\rightarrow\Pi^{aIJ}$ and the identification $4\gamma/(1+\s\gamma^2)\rightarrow a_1/a_2$ \cite{Celada}. What this transformation means is that the canonical momenta $\Pi^{aIJ}$ used in the canonical analysis of the CMPR action can be considered as the momenta $\stackrel{(\gamma)}{\Pi}\tensor{}{^{aIJ}}$ for which $\gamma$ satisfies the condition $4\gamma/(1+\s\gamma^2)=a_1/a_2$. Making these identifications, the constrains arising from the canonical analysis of the CMPR action become proportional (modulo terms proportional to the constraint $\Phi^{ab}$ introduced by $h_{ab}$ in the constraints $\mathcal{H}$ and $\Psi^{ab}$) to those resulting from the analysis of the action (\ref{BFact}) (see Ref.~\cite{Celada}). Thus, starting from the CMPR action instead of the action (\ref{BFact}) and bearing in mind the above identifications, one follows exactly the same procedure of sections \ref{SSolution} and \ref{SGauge}, and arrives at the same results contained there with the Immirzi parameter identified as above.

\section{Acknowledgments}

This work was supported in part by CONACYT, M\'{e}xico, Grant No. 167477-F.


\end{document}